\documentclass[10pt,a4paper,twoside,twocolumn]{article}
\usepackage[utf8]{inputenc}
\usepackage[T2A]{fontenc}
\usepackage[russian,english]{babel}
\usepackage{graphicx}
\usepackage{longtable}
\usepackage{rotating}
\usepackage[normalem]{ulem}
\usepackage{amsmath}
\usepackage{amssymb}
\usepackage{hyperref}
\usepackage{authblk}
\usepackage{isotope}
\usepackage[labelfont=bf, labelsep=period]{caption}
\usepackage{tabularx}
\usepackage{booktabs}
\usepackage[numbers, sort&compress]{natbib}
\usepackage{hyperref}

\author[1,2,3]{A. N. Ismailova \thanks{Email: ismailova@jinr.ru}}
\author[1]{Yu. L. Parfenova}
\author[1,4]{P. G. Sharov}
\author[1,2]{D. M. Janseitov}

\affil[1]{Flerov Laboratory of Nuclear Reactions, JINR, 20 Joliot Curie Str, Dubna, Russia}
\affil[2]{ Institute of Nuclear Physics, 1 Ibragimov Str, Almaty, Kazakhstan}
\affil[3]{Al-Farabi Kazakh National University, Almaty, Kazakhstan}
\affil[4]{Institute of Physics, Silesian University Opava, Czech Republic}			
\date{} 
\title{Model for the Glauber-type calculations of beam fragmentation at low energies\thanks{This work has been supported by the Science Committee of the Ministry of Science and Higher Education of the Republic of Kazakhstan (AP19678586)}}

\begin{document}
	\maketitle
	
	\begin{abstract}

  In the present paper momentum
  distributions of nuclei produced in the heavy ion beam fragmentation
  at the relatively low  energies  (below 100 \(A\cdot\)MeV) are
  studied. For this study, a new theoretical approach is developed on the basis of the Glauber model \cite{glauber1987high} modified for taking into
  account the energy and momentum conservation laws. In this approach,
  the longitudinal momentum of the most neutron rich
  nuclei,  \isotope[10]{Be}, \isotope[9]{Li}, \isotope[8]{He},
  produced in a few neutron removal reactions in the \isotope[11]{B}
  fragmentation in the Be target at a beam energy of 35 \(A\cdot\)MeV
  are calculated. The region of applicability
  of the new approach is discussed. This approach gives the
  asymmetric  longitudinal momentum distributions at low energies, and the
 asymmetry 
  is defined by the kinematical locus and geometry of the
  reaction (central of peripheral reactions).  We analyze the
  changes of the phase volume and the longitudinal momentum
  distributions with the beam energy and number
of the removed nucleons. The results of
  the calculations are compared to the parametrizations
  (see \cite{tarasov2004analysis}) widely
  used for estimates of nuclear production in fragmentation for
  planning of nuclear experiments.

\end{abstract}


\section{Introduction}
\label{sec:intro}

In-flight method of the ion beam production is widely used in experimental
nuclear studies. For effective functioning of the fragment
separator knowledge of the longitudinal momentum distributions of
fragments is of importance, especially for facilities
working at low energies ($E<100$ \(A\cdot\)MeV).  Thus, the longitudinal
emittance of the initial beam  2\%  leads to a 20\% increase after the
fragmentation, and the fragments are accepted into  the
secondary beam   within the narrow  region of the longitudinal
momentum distribution. The maximal yield of the fragments is
provided when the fragments are accepted in the vicinity of
the maximum in the momentum distribution, corresponding to the optimal conditions for the fragment acceptance.

For now, the nuclear fragmentation at the energies $E<100$ \(A\cdot\)MeV
is well studied, parametrizations of the fragment yields and
different standard approaches are suggested for calculations
of the fragmentation, giving a good description of the
experimental data at higher energies, $E>100$ \(A\cdot\)MeV, where
the Q-value of the reaction is negligibly small compared
to the beam energy. At the energies comparable to the
Q value, the regard for the energy and momentum conservation
laws is important.

For now, theoretical description of the differential  and inclusive
cross sections of the fragmentation at low energies, where
the transferred momentum and momentum of the beam are of the
same order of magnitude, is problematic. We suggest an approach
based on the Glauber model. It takes into account the
energy and momentum conservation
laws, and simultaneously keeps the simplicity and transparency of
the Glauber model.

In this approach, it is possible to get an analytical expression
for the amplitude of the process, and correct
it by the factor of the phase volume thus taking into account the
conservation laws.

In the present paper we apply the suggested approach for calculations
of the \isotope[11]{B} fragmentation in the Be target at the energy
35\(A\cdot\)MeV.
This is the lightest nucleus
where the few-proton removal leads to the production of the
\isotope[10]{Be},\isotope[9]{Li}, \isotope[8]{He} fragments. All these nuclear
fragments are thoroughly experimentally studied, providing
input parameters for
our approach. We analyze the changes of  the phase volume and the
longitudinal momentum distributions with the beam energy and mass number
of the fragments. We compare the momentum distributions to those obtained
with the systematics widely used for calculations of the fragmentation \cite{goldhaber1974statistical}, \cite{morrissey1989systematics}. We also discuss applicability of our
approach.

\section{Model description}
\label{sec:model}
We deduce the cross section and amplitude of the scattering process
within the scattering T-matrix formalism. In its terms, 
the differential cross section is expressed
as
\begin{equation}
  d\sigma=\dfrac{d\omega}{j}=|T|^2dV^{(n)}
\end{equation}
where \(T\) denotes the T-matrix and \(dV^{(n)}\)
is the phase volume of $n$-body system of fragments.

We suggest an approach where instead of the  T-matrix  we substitute the inelastic scattering amplitude multiplied by the Lorentz-invariant
phase volume.

\begin{figure}
	\centering
	\includegraphics[width=\linewidth]{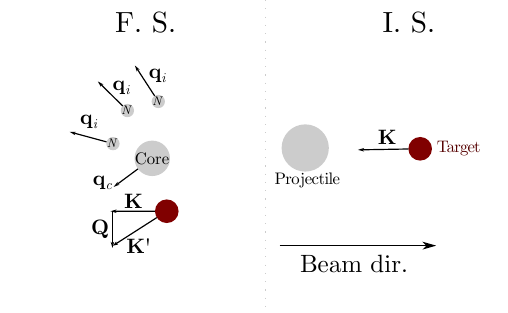}
	\caption{Kinematical scheme of the fragmentation
		reaction and kinematical variables used in the model.}
	\label{fig:kin-scm}
\end{figure}
To find the amplitude we consider the fragmentation of
the projectile (\(P\)) in the target
(\(T\)) (See Fig.~\ref{fig:kin-scm}) 
assuming that the projectile is composed by fragments, which are
relatively heavy core (\(C\))
and few nucleons. The initial state
of the projectile  is described by the wave function of the
relative motion of fragments with coordinates \(\mathbf{r}_i\).
We consider the problem in the projectile rest frame.
\(\mathbf{K}\) is the initial target momentum  and
\(\mathbf{K}'\) is that in the final state.
So we define \(\mathbf{Q}\) as the target transferred
momentum in the projectile rest frame,
\(\mathbf{q}_i\) are the projectile fragment momenta.
Also we define the longitudinal \(\mathbf{l}_i\) 
(with respect to \(\mathbf{K}\) )  and 
transversal \(\mathbf{t}_i\) components of \(\mathbf{r}_i\).

If we consider the initial wave function (WF) of the projectile as  
\[
  \Psi_I\equiv\Psi_I(\mathbf{r}_1,\dots,\mathbf{r}_n),
\]
where  \(n\) is the total number of fragments,
and the final state WF as
\[
  \Psi_F\equiv\Psi_F(\mathbf{r}_1,\dots,\mathbf{r}_n,\mathbf{q}_1,\dots,\mathbf{q}_n).
\]
Within the Glauber model,
neglecting the eclipse effect and
re-scattering we get the amplitude of the inelastic scattering as
\begin{equation}
  \mathcal{F}_{fi}=
  \sum\limits_{j=1}^nf_{j}(\mathbf{Q})
  \int  (\Psi_F)^* e^{i\mathbf{Q} \mathbf{r}_j} \Psi_I\prod_{i=1}^n d^3\mathbf{r}_i  .
  \label{eq:amp2}
\end{equation}
Here \(f_{j}\) is the fragment-projectile scattering amplitude.
Note, that it is assumed that \(\mathbf{Q}\) in \eqref{eq:amp2}
is in the plane orthogonal to the momentum \(\mathbf{K}\), and
\(\mathbf{q}_i\) should also be located in the same plane.

Our next assumption is the factorization of the WF
of the initial and final states
\[
\Psi_I\to \prod_{j=1}^N\Psi_{Ij}(\mathbf{r}_j)
\]
and
\[
  \Psi_F\to \prod_{j=1}^N\Psi_{Fj}(\mathbf{r}_j,\mathbf{q}_j)\to
  \prod_{j=1}^N\exp[i\mathbf{r}_j\mathbf{q}_j].
\]
Therefore we define functions
\begin{multline}
  F_j(\mathbf{Q},\mathbf{q}_1,\dots,\mathbf{q}_N)
  =\\
  \int{d^{3}}{\mathbf{r}_j}e^{i\mathbf{Q}\mathbf{r}_j}
  \Psi_{Fj}^{*}(\mathbf{r}_j,\mathbf{q}_j)\Psi_{Ij}(\mathbf{r}_j)\times\\
  \prod_{k\ne j} \int d^{3}{\mathbf{r}_k}
  \Psi_{Fk}^{*}(\mathbf{r}_k,\mathbf{q}_k)\Psi_{Ik}(\mathbf{r}_k).
\end{multline}
After approximation of the final state WFs by the plane waves we get
\begin{multline}
  F_j(\mathbf{Q},\mathbf{q}_1,\dots,\mathbf{q}_N)=\\
  \int{d^{3}}{\mathbf{r}_j}e^{-i(\mathbf{q}_j-\mathbf{Q})\mathbf{r}_j}
  \Psi_{Ij}(\mathbf{r}_j)\times\\
  \prod_{k\ne j} \int{d^{3}}{\mathbf{r}_k}e^{-i\mathbf{q}_k\mathbf{r}_j}
  \Psi_{Ik}(\mathbf{r}_k)=\\
  F_j(\mathbf{q}_j-\mathbf{Q})\prod_{k\ne j}F_k(\mathbf{q}_k).
  \label{eq:formfactor}
\end{multline}

After substitution of \eqref{eq:formfactor} into \eqref{eq:amp2} we obtain the
amplitude as
\[
  \mathcal{F}_{fi}=
  \sum\limits_{j=1}^Nf_{j}(\mathbf{Q}) F_j(\mathbf{q}_j-\mathbf{Q})\prod_{k\ne j}F_k(\mathbf{q}_k).
\]
We will show that \(f_C\gg f_N\)
so we can leave only one term in the expression.

Thus, finally,
the amplitude of the inelastic scattering takes the form
\begin{equation}
  \mathcal{F}_{fi}\approx
  f_{c}(\mathbf{Q}) F_C(\mathbf{q}_C-\mathbf{Q}) \prod_{k} F_N(\mathbf{q}_k),
  \label{eq:amp3}
\end{equation}

Note here that $ F_C(\mathbf{q}_C)$ is the form factor determined
by the nucleus size. Thus, the inelastic scattering amplitude is defined by the elastic scattering amplitude of the core $f_{C}(\mathbf{Q})$. The interaction with nucleons enters the
form factors $F_N(\mathbf{q}_k)$. The obtaining of this expression
is given in \cite{glauber1987high}.

We approximate the functions \(F_C\) and \(F_N\)  by the oscillator function
in the  momentum space as
\begin{equation}
  F_j(q)=  \sqrt[4]{\frac{32}{27}\pi \langle r_j\rangle^{6}
    e^{-\frac{4}{3}q^2\langle r_j\rangle^2}};
    \label{eq:ffexp}
\end{equation}
where \(\langle r_j\rangle\) is root mean squared radius (RMS) of the
fragment WF.

The elastic scattering amplitude $ f_{C}(\mathbf{Q})$ is calculated
in the Glauber model using the corresponding core -target profile
function $S_C$ as
$$ f_{C}(\mathbf{Q}) = <\Psi _F |S_C| \Psi_I> $$
where the profile function  is obtained  using the optical
model potential

\begin{equation}
	S_{C}(b) = \exp\left[ -\frac{i}{\hbar v} \int_{-\infty}^{\infty} dz \, V_{C T} \left( \sqrt{b^2 + z^2} \right) \right],
\end{equation}
 where \( V_{C}(r) \) is
the optical potential describing the core-target
interaction.
\( b \) is the impact parameter of the center of mass
of the core (see, for example,\cite{hencken1996, parfenova2000}
and references therein).$v$ is the beam velocity, and $z$ is the  coordinate
along the beam axis.

There are a lot of parametrizations of the optical model parameters
available  in the literature. In our case, when we develop the
approach for a very wide projectile energy range, we use the
standard parametrization of
the nucleon-nucleon interaction potential with the parameters from
\cite{Charagi1990, Ray1979}, which is valid for the incident
energy range from 10 to 2000
\(A\cdot\)MeV.

\subsection{Folding Model of the Optical Potential}

The nucleon-nucleon potential describes the interaction of each nucleon composing the nucleus with each nucleon of the target.
To calculate the optical potential of the core-target interaction we find the folding  \cite{varner1991physrep} of the potential with the
nucleon densities. 

The core-target interaction potential in the folding
model can be expressed as follows:

\begin{equation}
	V_{C}(\mathbf{r}) = \int A_{C}\rho_{C}(\mathbf{r'}) \overline{V_{CT}}(|\mathbf{r} - \mathbf{r'}|) \, d\mathbf{r'}.
\end{equation}
where 
\begin{equation}
	\overline{V_{C T}}(|\mathbf{r} - \mathbf{r'}|) = -\frac{i}{2}\hbar v A_T \rho_T(|\mathbf{r} - \mathbf{r'}|) \overline{\sigma_{NN}}
\end{equation}
The density distribution of the interacting nuclear systems is approximated by a Gaussian distribution \cite{parfenova2002}:

\begin{equation}
	\rho(x) = \rho_0 \exp(-\alpha x^2),
\end{equation}
where \( \alpha = \left[\frac{2}{3} \langle r^2 \rangle \right]^{-1} \) is the density distribution parameter related to RMS of the nucleus \( \langle r^2 \rangle^{1/2} \), and \( \rho(x) \) is normalized to unity.
\( \overline{\sigma}_{NN} \) is the nucleon-nucleon cross-section averaged over the number of neutrons and protons involved in the interaction ( for more
details, see \cite{Charagi1990}, \cite{Ray1979}).

\begin{figure}
  \centering
  \includegraphics[width=\linewidth]{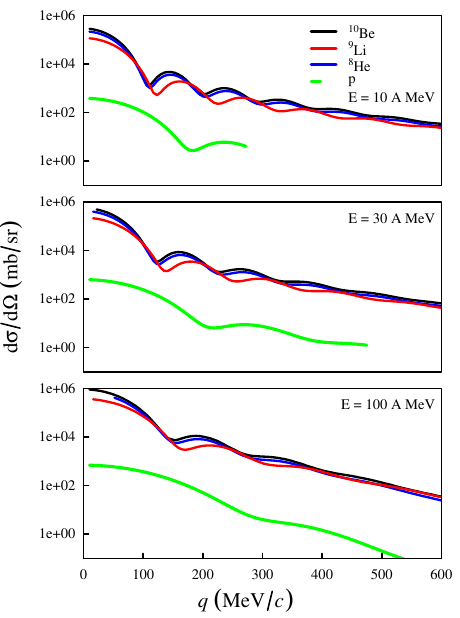}
  \caption{\label{fig:el-sc}
    Differential elastic scattering as a function of the transferred momenta
    \(q\) for scattering of protons and heavy ions
    (\isotope[10]{Be}, \isotope[9]{Li}, and \isotope[8]{He} ) on the
    \isotope[9]{Be} target for different projectile energies.
    (see the legends on the panels.)   }
\end{figure}
\section{Results and discussion}
\label{sec:resanddis}
As mentioned above, we consider three reactions
(for different beam energies) 
\begin{gather*}
  \isotope[11]{B}+  \isotope[9]{Be}\to  \isotope[9]{Be} +  \isotope[10]{Be} +p,\\
  \isotope[11]{B}+  \isotope[9]{Be}\to  \isotope[9]{Be} +  \isotope[9]{Li} +2p,\\
  \isotope[11]{B}+  \isotope[9]{Be}\to  \isotope[9]{Be} +  \isotope[8]{He} +3p,\\
\end{gather*}
which ``illustrate'' the path towards the neutron dripline.

In our model, we consider the projectile nucleus \isotope[11]{B}
as a system of pre-formed heavy cluster and valence protons.
The structure of \isotope[11]{B}
is described by the corresponding form factor $ F_C(\mathbf{q}_C)$
in (\ref{eq:ffexp})
determined by the RMS distance of the core inside \isotope[11]{B}.
In our calculations, we used the proton RMS distance $
 \langle r_p \rangle = 2.43~(\mathrm{fm})$
taken in accordance with the standard systematics \cite{Marinova}.
As the first approach one can obtain \(\langle r_C \rangle\)
supposing the core to be a particle.
The values of \(\langle r_C \rangle\) are presented in the
Table~\ref{tab:wf-par}.
Since the core WF inside \isotope[11]{B} may be more complicated,
we vary \(\langle r_C \rangle\) in a wide range
(see also the Table~\ref{tab:wf-par}) to demonstrate
sensitivity of our model to this parameter.
\begin{table}[t]
  \centering
  \caption{\label{tab:wf-par}
    The RMS radii of heavy fragment used in the calculations.
    \(\langle r_c \rangle\) is the default value;
    \(\min\langle r_c \rangle\) is the radii for case of
    ``more central'' reaction;
    \(\max\langle r_c \rangle\) is the radii for case of
    ``more peripheral''  reaction.
    (all the values are given in fm,)
  }
  \begin{tabularx}{\linewidth}{Xccc}
    \toprule
    Fragment  &\(\langle r_c \rangle\) &\(\min\langle r_c \rangle\) & \(\max\langle r_c \rangle\) \\
    \midrule
    \textsuperscript{10}Be &  0.2426160 & 0.1213080 & 0.4852321 \\
    \textsuperscript{9}Li &   0.4852321 & 0.2426160 & 0.9704641 \\
    \textsuperscript{8}He &   0.7278481 & 0.3639241 & 1.4556962 \\ 
    \bottomrule
  \end{tabularx}

\end{table}

For the calculations of the reaction rate \eqref{eq:amp3}
the Monte Carlo method is used.

\begin{figure*}[!p]
  \centering
  \includegraphics[width=0.8\textwidth]{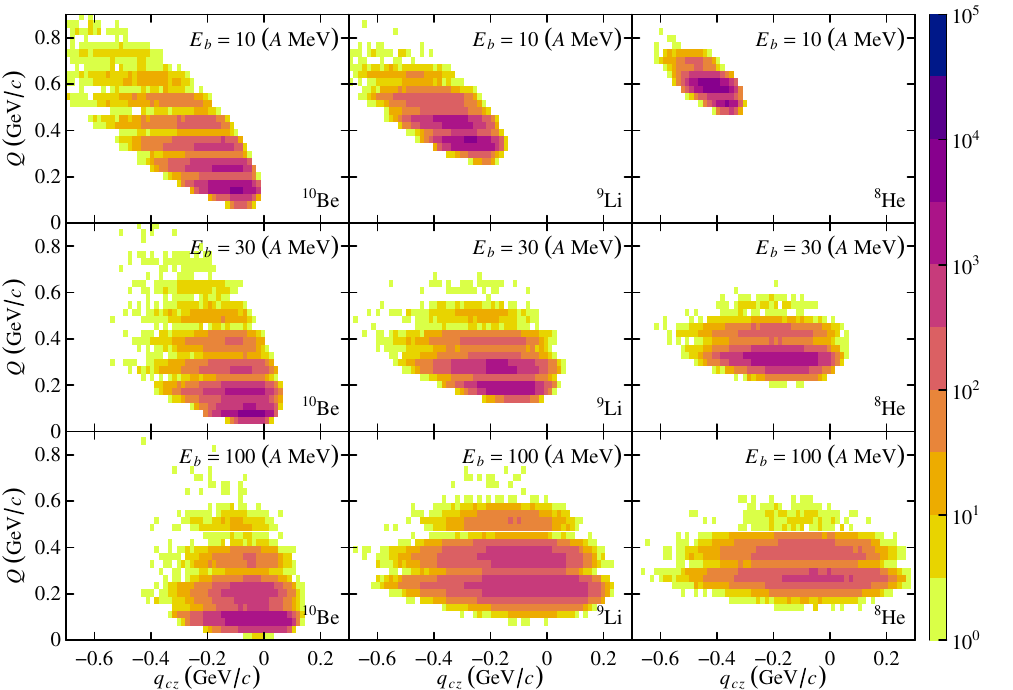}
  \caption{\label{fig:Q-qcz}
     Correlation of the transferred momentum of target  \(Q\)
    and longitudinal part \(q_{cz}\)  of the core momentum
    in the projectile rest frame 
    for different fragments (shown in bottom right panel corner)
    and different \isotope[11]{B} beam energies
    (shown in top right panel corner).
  }
\end{figure*}

\begin{figure*}[!p]
  \centering
  \includegraphics[width=0.8\textwidth]{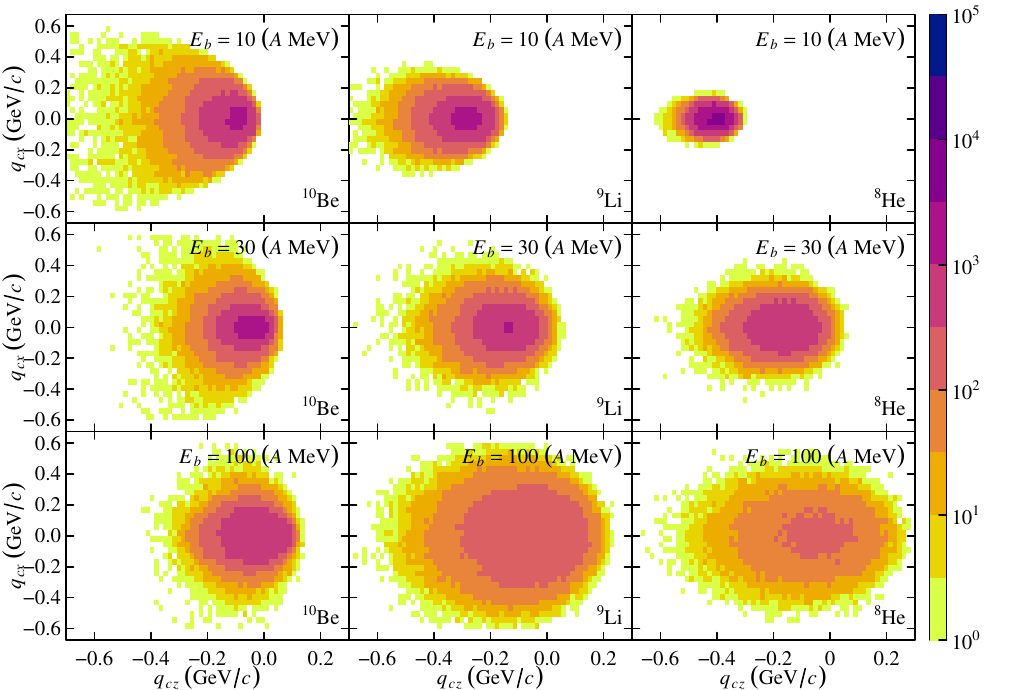}
  \caption{\label{fig:qcx-qcz}
    Correlation of the transversal \(q_{cx}\)  and
    longitudinal \(q_{cz}\) parts
    of the core momentum in the projectile rest frame.
    The layout is the same as in Figure~\ref{fig:Q-qcz}.
  }
\end{figure*}

The Figs. \ref{fig:Q-qcz} and \ref{fig:qcx-qcz} show the correlation
plots for the target transferred momentum (\(Q\)) vs. projection of 
the core momentum in the projectile rest frame where the $z$
axis coincides with the beam direction,
and similar dependence of 
the two projections of the core momentum in the projectile rest frame
( \(q_{cx}\)  and  \(q_{cz}\)).

The kinematical loci of the fragments can be easily recognized in
Figs. \ref{fig:Q-qcz} and \ref{fig:qcx-qcz}.
Considering these graphs, few moments should be pointed out which are
as follows:
(i) for the low beam energies, 10 and 30 MeV, the
variables \(Q\) and \(q_{cz}\)
are strongly correlated; this correlation becomes weaker with
increase in the number of fragments;
(ii) the loci of \(q_{cz}\) are asymmetric relative to the beam speed point,
so there is a tendency of ``slowing down'' of the fragment at low energies
related with the reaction kinematics.





In the calculations of the  elastic scattering,
we expected that the dominant part of the
elastic scattering cross section corresponds to the transferred momentum
lower than that of the first diffraction minimum (See Fig. \ref{fig:el-sc}).
Note, that kinematic locus in this case contains the non-zeroth
transferred momentum $Q$ since the part of the beam energy is spent for
the nucleon knock-out. Therefore, the case of the Serber model \cite{serber1963theory}, which is valid for the limit $Q=0$ MeV/c, is not realized due to the kinematical
locus. In this case the fragments are slower than the beam nuclei.

This result shows necessity of the momentum and energy conservation
laws at low energies. From this viewpoint, the models of
the sudden removal with $Q=0$, such as the Serber model and
the traditional eikonal approximation of the Glauber model, give
only qualitative description, and don't reproduce the position of
the momentum distributions, while their widths are close to
those obtained in our approach.


\begin{figure}
  \centering
  \includegraphics[width=\linewidth]{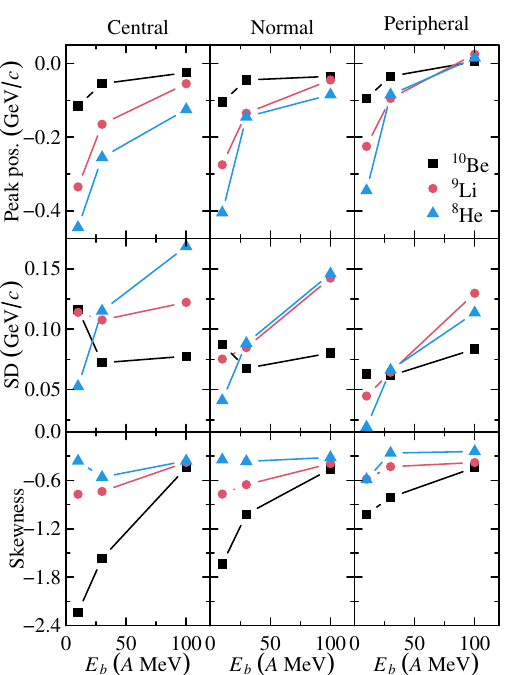}
  \caption{\label{fig:rcvar}
    The mode (peak position), standard variation (SD),
    and skewness for different values of \(\langle r_C\rangle\)
    (see the Table~\ref{tab:wf-par}).
  }
\end{figure}

Sensitivity of our calculation to the \(\langle r_C\rangle\) parameter
is illustrated by Figs.  \ref{fig:rcvar} and \ref{eq:ffexp}.
The variation of the results of the calculations with this parameter
can be explained as follows: in the case of the
small values of \(\langle r_C\rangle\) for the particular
reaction channel, the central collisions dominates, while the 
large values of \(\langle r_C\rangle\) the peripheral process becomes
essential.
To demonstrate these effects we performed the calculations with
different values of \(\langle r_C\rangle\), the minimal, default, and
maximal ones presented in the Table~\ref{tab:wf-par}.

Figure~\ref{fig:rcvar} shows the changes of the mode (peak position),
standard variation,
and skewness  of the
\(q_{cz}\) distribution with the  \(\langle r_c\rangle\) value.
The qualitative behavior of the \(q_{cz}\) distribution remains the same.
For the peripheral case, one can see ``speed-up'' of the fragment,
(for the high energy \(q_{cz}\) is faster then the beam velocity point)
and narrowing of the \(q_{cz}\) distribution.
For the central case, the \(q_{cz}\) distribution
becomes wider and more asymmetric.

\begin{figure}
  \centering
  \includegraphics[width=\linewidth]{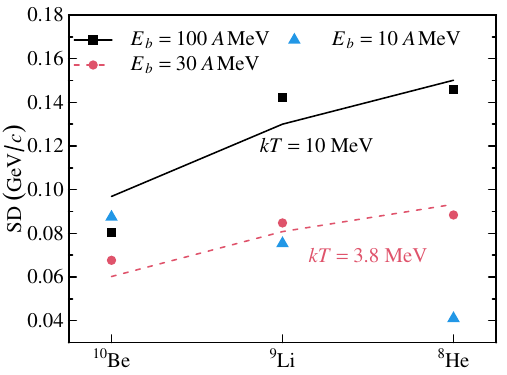}
  \caption{\label{fig:sdpar}
    Comparison of standard deviation (SD) of \(q_{cz}\) obtained in our
    calculations (points) and that obtained in the model 
    \cite{goldhaber1974statistical} (lines).
    The values of \(kT\) (see \cite{goldhaber1974statistical} )
    for the beam energy
    \(E_b=100\)~\(A\)~MeV and \(E_b=30\)~\(A\)~MeV
    are presented in the panels.
  }
\end{figure}

Finally in Figure~\ref{fig:sdpar}  we compare the calculated width of the
longitudinal momentum distributions of the fragment with the
widely used Goldhaber  model  \cite{goldhaber1974statistical}.
For the high beam energy     \(E_b=100\)~\(A\)~MeV and \(E_b=30\)~\(A\)~MeV
our results are in the reasonable agreement with \cite{goldhaber1974statistical}.
For the low beam energy \(E_b=10\)~\(A\)~MeV the fragment mass number
dependence of the
widths of the longitudinal momentum distributions in our calculations
qualitatively differs from that obtained in \cite{goldhaber1974statistical}.

\section{Summary}
\label{sec:summary}

One of the most traditional ways of the exotic (neutron or proton-rich) nuclei
production is the fragmentation of the beam nuclei. To provide the effective production of a certain isotope the preliminary calculations are needed. At the energies $E<100$ \(A\cdot\)MeV, traditional sudden removal approaches are not precise. 

We suggest the approach based on the Glauber model modified for regard for the energy and momentum conservation laws. For calculations of the amplitude of the fragmentation reaction, the Lorentz-invariant phase space is introduced. 
The analytical solution exists for the case of the elastic scattering. The parameters of the approach , such as rms radii of a nucleus  and the related rms distance between core and valence nucleons in the exotic nuclei, the optical model parameters, etc., are well known, and can be found in literature on the experimental data and parametrizations.

With the example of the \isotope[11]{B} fragmentation we have calculated the longitudinal momentum distributions of the \isotope[10]{Be}, \isotope[9]{Li},
and \isotope[8]{He} fragments. The relation between the longitudinal momentum distribution and number of removed nucleons is studied, as well as the fragment mass and beam energy dependence are analyzed. It is found that the heavier the target, the better the precision of the approach. We also see, that for the energies below 100 MeV the fragments are slower than the beam nuclei, that is defined by the kinematic locus, while for the high energies (>100 A MeV)  the fragment may appear to have both higher or lower speed depending on geometry of the reaction. In particular, more peripheral reactions lead to faster fragments, and less peripheral (central) reactions provide the slower fragments with respect to the beam velocity.

The regard for the energy and momentum conservation laws leads to essential changes in the shape of the longitudinal momentum distributions, showing asymmetry, where the low momentum tail of the distribution is formed due to the large transferred momentum. Thus, compared to the Glauber model our approach describes the fragmentation in a wider range of momentum transfer. The Glauber provides a good description within the momentum transfer 100-150 MeV/c. The applicability of our approach for the momentum transfer beyond the first diffraction minima needs more detailed studies. In the present paper we show, that the transfer on nucleons can proceed with the momentum transfer essentially exceeding that corresponding to the first diffraction minimum. The approach allows to reproduce the wide momentum distributions in this case.

The comparison with calculations within  other models and parametrizations shows
that at the energies less then 100 \(A\cdot\)MeV the kinematical loci and
energy-momentum conservation laws 
should necessarily be taken into account while planning the experiment
and finding the optimal conditions for the fragment production.

\bibliographystyle{unsrt}
\bibliography{simple-model}

\end{document}